\documentclass[prl,aps,showpacs,twocolumn]{revtex4}

\usepackage{graphicx}
\usepackage{braket}
\usepackage{bm}

\begin{document}
	
\def \o2{O$_2$}
\def \us{$\mu$s}

\title{Trapping of molecular Oxygen together with Lithium atoms}

\author{Nitzan Akerman, Michael Karpov, Yair Segev, Natan Bibelnik, Julia Narevicius and Edvardas Narevicius.}
\affiliation{Chemical Physics Department, Weizmann Institute of Science.}

\date{\today}

\begin{abstract}
We demonstrate simultaneous deceleration and trapping of a cold atomic and molecular mixture. This is the first step towards studies of cold atom-molecule collisions at low temperatures as well as application of sympathetic cooling.
Both atoms and molecules are cooled in a supersonic expansion and are loaded into a moving magnetic trap which brings them to rest via the Zeeman interaction from an initial velocity of 375 m/s. We use a beam seeded with molecular Oxygen, and entrain it with Lithium atoms by laser ablation prior to deceleration. The deceleration ends with loading of the mixture into a static quadrupole trap, which is generated by two permanent magnets. We estimate $10^9$ trapped \o2 molecules and $10^5$ Li atoms with background pressure limited lifetime on the order of 1 second. With further improvements to Lithium entrainment we expect that  sympathetic cooling of molecules is within reach.
\end{abstract}

\maketitle
During the last decades trapping and cooling of atoms became a workhorse of atomic physics with countless experiments where atoms routinely cooled down to nano-Kelvin temperatures. On the other hand, cooling and trapping of molecules remains challenging. Several methods demonstrated molecular trapping and cooling including assembly of ultracold diatomic molecules from laser-cooled alkali-metal atoms \cite{miller1993photoassociation,ni2008high,park2015ultracold,takekoshi2014ultracold}, direct laser cooling and magneto-optical trapping of molecules with favorable vibrational transitions that allow scattering a large number of photons \cite{Norrgard2016Submillikelvin,zhelyazkova2014laser}. Sisyphus molecular cooling has been demonstrated in the case of electrostatically trapped molecules \cite{prehn2016optoelectrical}.
It is also possible to cool molecules without relying on laser transitions via collisions with cold buffer gas \cite{hutzler2012buffer}. Cold molecules can be extracted from a cryogenic buffer gas cell in hydrodynamically enhanced flow, producing  an intense and versatile source which can be used as a starting point in other experiments \cite{patterson2007bright}. A similar direct and general molecular cooling method is based on collisions that occur during adiabatic expansion of high pressure gas into vacuum.  Atomic and molecular beams that are formed by such expansion have been successfully decelerated using inhomogeneous magnetic and electric fields\cite{lemeshko2013manipulation}. Subsequent molecular trapping of Stark decelerated beams has been demonstrated using electric and magnetic traps \cite{bethlem2000electrostatic,van2005deceleration,stuhl2012evaporative}. In a similar fashion paramagnetic atomic or molecular beams have been trapped following Zeeman deceleration \cite{vanhaecke2007multistage,weinstein1998magnetic,liu2016magnetic}. 

Far less progress has been made in generation of cold mixtures of atoms and molecules, even though it opens many possibilities in both physics and chemistry. An immediate advantage that atom-molecule co-trapping offers is the orders of magnitude longer interrogation times compared to molecular crossed beam methods. This will enable the study of cold chemistry for especially slow processes. Particularly, inelastic collisions in such a setup have been already successfully studied by Parazzoli et al. \cite{parazzoli2011large} and the upper limit on reactive collisions between N and NH has been placed by Campbell et al. \cite{campbell2007magnetic}. In addition, as in the production of cold alkali molecules, photoassociation or Feschbach resonances can be used to construct polyatomic cold molecules. 
One of the most exciting opportunities that cold atom-molecule mixtures enables is the application of sympathetic cooling, where cold atoms that are amenable to laser cooling can be used to collisionally cool molecules. Prospects of such a cooling scheme strongly depend on the collisional properties of atoms and molecules with several candidates identified \cite{wallis2009production} and cooling mechanism studied in detail \cite{lim2015modeling}. A general guide for a successful application of sympathetic cooling is weak anisotropy in the interaction potential and low reduced mass of interacting particles that helps to suppress inelastic scattering channels by increasing the centrifugal barrier height in the exit channel.

In order to achieve the highest possible trapped molecular density, we choose molecular Oxygen that can be easily seeded in a supersonic expansion and decelerated via Zeeman interaction, with mass to magnetic moment ratio of $\sim$16 amu/Bohr magneton. Deceleration of \o2 molecules by pulsed magnetic fields has been demonstrated  \cite{narevicius2008molecular,wiederkehr2012velocity}. Recently, Liu et al. have also reported short confinement, on the order of 600 \us, of the molecular packet in an electromagnetic trap \cite{liu2015one}.

\begin{figure*}[ht!]
	\centerline{\includegraphics[width=16 cm ]{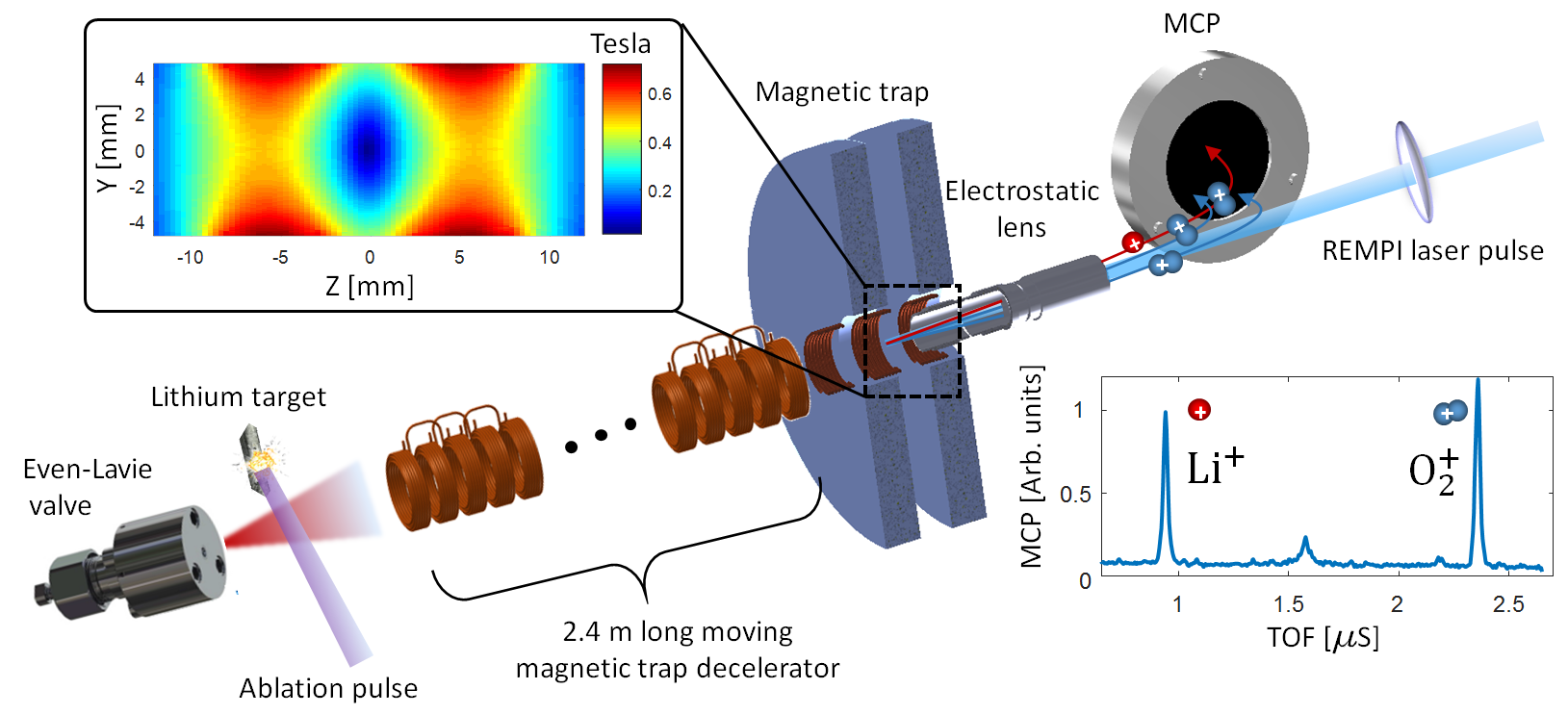}}
	\caption{Schematic diagram of the experiment. A pulsed supersonic beam of \o2 seeded in Kr is produced by an Even-Lavie valve. Li atoms can be entrained into the beam by laser ablation of a solid target placed near the valve. Low field seeking states are decelerated by a 2.4  meter long moving trap decelerator and then trapped in a permanent magnet quadruple trap. Both \o2 molecules (REMPI) and Li atoms (single photon) are ionized by a 225 nm laser pulse. The ions are then extracted by an electrostatic lens towards an MCP detector. The right inset shows a TOF trace with two prominent peaks which correspond to the two products. Left inset shows the calculated magnetic field magnitude of the permanent trap.}				
	\label{setup}
\end{figure*}

Since most of the atoms are paramagnetic in the ground or long lived metastable state many are well suited for Zeeman deceleration. In a previous work, we have demonstrated co-deceleration of metastable Argon atoms together with molecular Oxygen \cite{akerman2015simultaneous}. In that experiment Oxygen was cooled in a molecular expansion and metastable Argon was generated by electric discharge from the carrier Argon gas. 
Here, we go beyond deceleration and demonstrate how the decelerated molecular ensemble can be transferred into a permanent trap in order to open the possibility of sympathetic cooling. Furthermore, we have co-trapped Lithium atoms, by entraining them into the beam prior to deceleration. Importantly metastable Argon is unstable with the Penning ionization process taking place with both molecular Oxygen and Ar*. In contrast, the ground state Li and \o2 reaction is endothermic and does not occur at our trapping temperatures. Moreover, Li has the advantages of lower reduced mass and is highly suitable for laser cooling. Throughout the deceleration and subsequent trapping the atoms and molecules are confined in a 3D trap, leading to small losses  during both the deceleration and transfer into the permanent trap. We estimate $10^9$ trapped \o2 molecules (with density of n=$10^{10}$cm$^{-3}$) at a temperature of 300 mK together with $10^5$ Lithium atoms in a permanent magnetic quadrupole trap.
Our results here provide a pathway to further implementation of molecular cooling by forced evaporation and sympathetic cooling.   

Our experimental apparatus is presented in Figure 1. A pulsed beam of \o2 is produced by expanding a mixture of \o2 and Kr with stagnation pressure of about 10 bar into vacuum using an Even-Lavie valve \cite{even2014pulsed} . The valve is cooled to a temperature of 165 K in order to reduce the mean initial velocity of the beam to below 400 m/s, from which it can be decelerated to a stop. The velocity spread of the beam is $\pm$ 25 m/s, which corresponds to translational temperature of $\approx 3$ K. In other works \cite{wiederkehr2012velocity} it was found that the rotational temperature of \o2 seeded in Kr was around 5 K, which means that only the N=1 manifold (lowest state for \o2) is occupied.

At a distance of about 15 cm the cold supersonic beam enters a moving trap Zeeman decelerator. The working principles of our decelerator are given elsewhere \cite{lavert2011moving,akerman2015simultaneous}. Briefly, it consists of 480 spatially overlapping quadrupole traps and spans over 2.4 meters long. The traps are activated sequentially in a temporally overlapping manner, where each pulse follows a half sine shape and each trap is activated at the peak current of the preceding trap. The instantaneous velocity of the decelerated beam is controlled by the width of each current pulse. All the pulses are generated by ten driver-modules, which are real-time configurable LC circuits capable of delivering up to 600 A with variable pulse duration ranging form 20 to 500 \us. These currents can generate magnetic fields as high as 0.8 Tesla along the trap symmetry axis, which corresponds to a trap depth of about 300 mK in the transverse direction and 400 mK in the longitudinal for \o2 at a deceleration value of 35,000 m/s$^2$. Using a computer-controlled pulsed sequence we can set the final beam velocity to anywhere between 450 m/s and 20 m/s, which is slow enough for loading into a static magnetic trap.

Once the molecules and the atoms are slowed to low enough velocities, they can be trapped in a static trap. Generating a magnetic field on the order of 1 Tesla by running a current in a coil for millisecond time scales is fairly straightforward. Extending it to seconds becomes a much more difficult problem, as one needs to deal with a significant amount of heat dissipation. To circumvent this problem we use a static magnetic trap, that is formed by two Neodymium Iron Boron permanent magnets. Each magnet has an outer diameter of 69 mm, a central bore with a diameter of 12 mm and a width of 6 mm. The on-axis magnetic field peak of a single magnet has a magnitude of about 0.5 Tesla. The two magnets are aligned in opposite directions and separated by 1 cm (center to center) to form a quadrupole trap with a longitudinal depth of 0.5K and radial depth of 0.3K  for \o2 molecules. Although particles in such a trap are subject to non-adiabatic spin flip (Majorana) losses near the trap center, this part of the phase space is negligible for our experimental parameters and therefore does not result in significant loss.

\begin{figure}[ht!]
	\centering
	\includegraphics[width=9 cm]{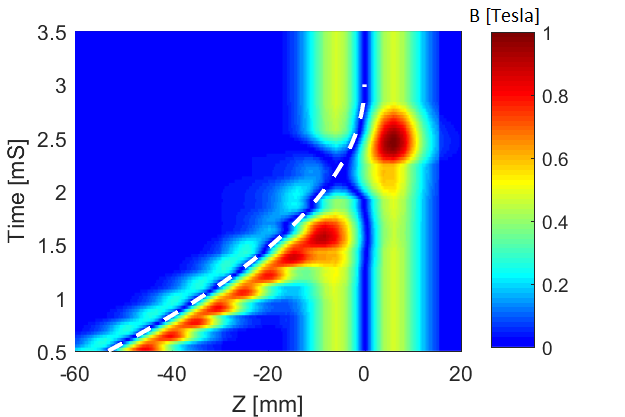}
	\caption{Loading sequence of the static magnetic trap, showing the magnetic field magnitude along the Z axis as a function of time. Additional coils are used to cancel the front permanent magnet's field at the right time to allow the molecules to enter the static trap. The minimum of the magnetic field closely follows a trajectory of constant deceleration (dashed white line) that reaches zero velocity at the static trap's center.}
	\label{loadingSeq}
\end{figure}

During the loading process into the permanent magnets trap the front barrier needs to be momentarily eliminated, in order to let the slow molecular beam enter into the center of the trapping region. This is achieved by using three extra coils, which adiabatically guide the molecules from the last decelerator trap into the static trap. The first coil generates a quadrupole trap together with the front magnet, bringing the molecules as close to it as possible. Then a second coil is used to cancel the magnetic field of the front magnet, letting the molecular beam to enter the trapping region. A third coil is used to increase the back magnet's field in order to bring the molecules to a stop close to the trap center. Figure \ref{loadingSeq} shows the calculated magnetic field magnitude along the symmetry axis as a function of time during the loading process. The dashed white line is a parabolic curve, which illustrates a trajectory of constant deceleration to zero velocity at the static trap center, which coincides well with the generated magnetic field minimum.       

Unfortunately \o2 does not possess a dipole transition from the ground state in the accessible optical spectrum that can be used for detection and manipulation. Moreover, those exited states which do exist are in the extreme UV and undergo fast pre-dissociation. Therefore, for detection of the \o2  molecules we use a resonance enhanced multi-photon ionization (REMPI) process, where the produced ions are then measured with a multi-channel plate (MCP) detector. Here we use a 2+1 REMPI process at 225 nm \cite{yokelson1992identification}. The two-photon transition excites the 3d$\pi^3\Sigma_1^-(\nu'=2)\leftarrow$ X$^3\Sigma_g^-(\nu''=0)$ intermediate level and the third photon ionizes it into the continuum.  
In order to minimize any free flight of the molecules, the static trap magnets are mounted on the same 1 cm diameter vacuum tube of the decelerator. As a result, the detection region has a single optical axis in the counter-direction of the beam. We use an electrostatic lens to extract the ions from the grounded vacuum tube and direct them towards the MCP. The lens consists of three metal tubes which are inserted into the vacuum tube and their voltages can be controlled externally. The right inset of Fig. 1 shows a typical time of flight (TOF) trace measured by the MCP, where the two dominant peaks correspond to the different ionic outcomes of the REMPI pulse O$_2^+$ and Li$^+$ (the small peak in the middle is a O$^+$). 

\begin{figure}[ht!]
	\centering
	\includegraphics[width=9 cm]{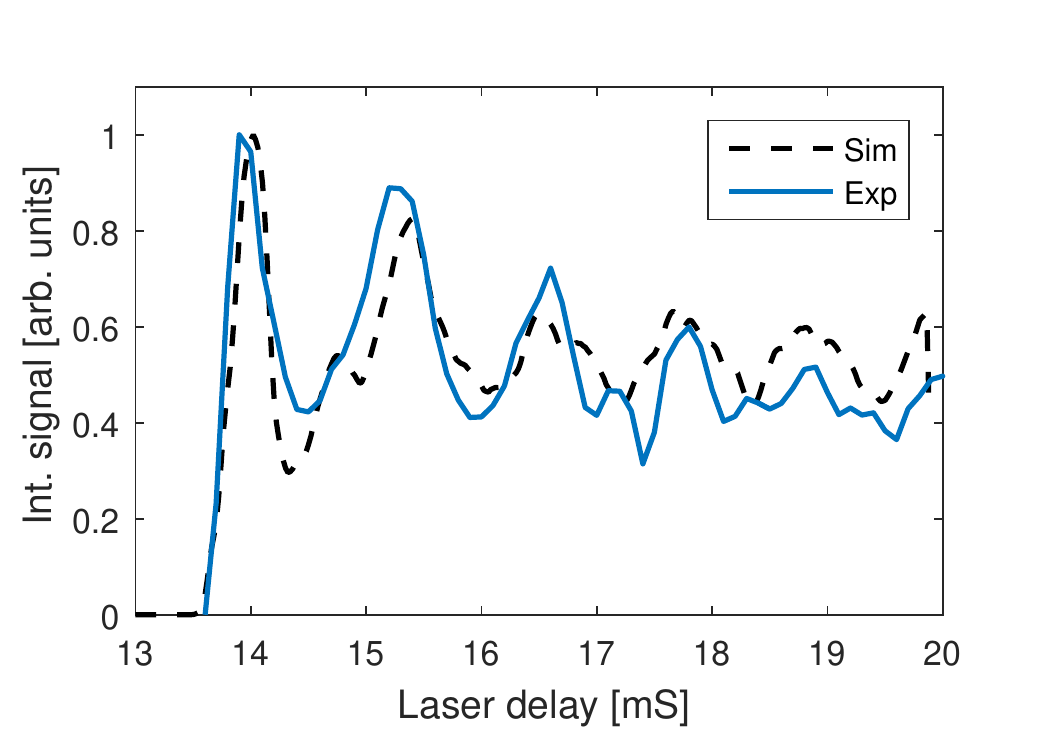}
	\caption{Measured and simulated loading dynamics of the static magnetic trap. Measured REMPI signal (solid black) from the static trap center as a function of time. The dashed blue line is the result of a Monte-Carlo simulation of 4000 molecules showing density at the trap center (average on volume of 10 mm$^3$). }
	\label{loadingO2}
\end{figure}

Figure \ref{loadingO2} presents the measured \o2 REMPI signal during the first few milliseconds of trapping, along with results from a simulation of the trap dynamics. The highest initial peak comes from the decelerated \o2 packet crossing the center of the trap, followed by few additional density oscillations with a period of around 1 ms, which settle to about 50 \% of the peak density after a few oscillations. These oscillations indicate a non-perfect adiabatic loading with a small residual mean velocity. This is expected, because reaching zero velocity in a perfect adiabatic sequence would take infinite time. In our case, a non-perfect adiabatic loading compromises the resulting phase space through the time it takes to enter the trap (canceling the field of the front magnet). This is further evident in the 50 \% loading efficiency. 
We simulate the loading process with 4000 molecules and get a good quantitative agreement with the experiment.      

The Li atoms are entrained into the beam by ablation of a solid Lithium target, placed near the nozzle on a wobble stick. This is achieved by using pulsed laser (Quanta Ray Indi) with pulses of few 10s of mJ at 355 nm, focused on the target. In this way we have entrained Li atoms into the supersonic beam and trapped them together with the \o2 molecules in the permanent magnetic trap. The 225 nm laser pulses have sufficient energy to ionize the Li atoms together with the \o2 molecules, such that the extraction and detection procedures for the two species are the same and are performed simultaneously. 

\begin{figure}[ht!]
	\centering
	\includegraphics[width=9 cm]{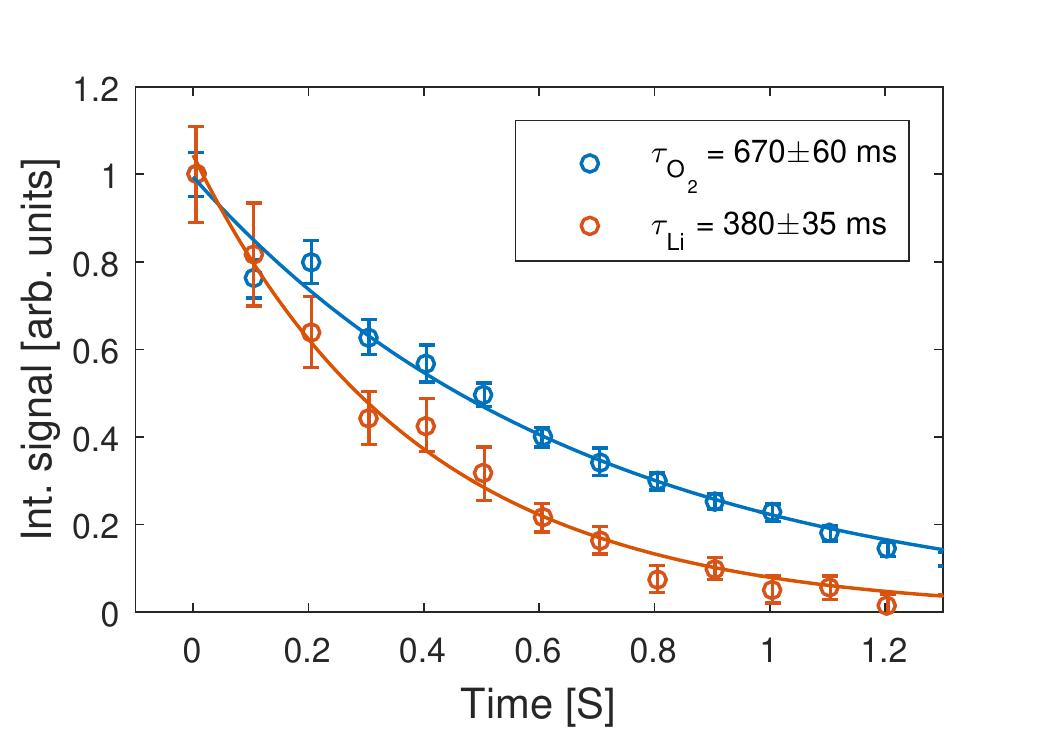}
	\caption{Measured lifetime of the trapped \o2 molecules and Li atoms in the permanent trap. From fitting the data to a decaying exponent (solid line) we infer a 1/e lifetime of $\tau_{O_2}=670\pm60$ ms for the Oxygen and $\tau_{Li}=380\pm35$ for the Lithium.}
	\label{O2_Li_2}
\end{figure}

Figure \ref{O2_Li_2} shows the \o2 and Li REMPI signals as a function of time from the moment the trap was loaded. Each data point is an average of 40 repetitions. From fitting the measured results to an exponential decay we infer a trap lifetime of $670\pm60$ msec for the \o2 molecules and $380\pm35$ msec for the Li atoms. We assume our lifetime is limited by the background pressure, which in this case implies a few $10^{-8}$ Torr \cite{arpornthip2012vacuum}. Although our vacuum gauge reading is 10 times smaller, such a pressure gradient between the gauge and the trap is reasonable due to geometry and vacuum conductance. The difference in lifetime is consistent with the ratio of 1.8 obtained by calculating the collision rates of Li-H$_2$ and \o2-H$_2$ systems based on van der Waals dispersion coefficients \cite{bali1999quantum,rijks1989frequency}. In order to verify that the Lithium lifetime is limited by the background and not by inelastic collisions with the Oxygen molecules, we have performed the same measurement using a beam of pure Kr, in which the Li atoms have been entrained. The Lithium lifetime obtained from this measurement is almost the same, which supports our assumption.

Estimating the absolute number of trapped molecules from the REMPI signal is a difficult task, as the process efficiency is not well known and strongly depends on the laser beam intensity and shape. Therefore, our best estimation relies on a residual gas analyzer (SRS model RGA100) measurement. To calibrate the RGA sensitivity to \o2 we flooded the vacuum chamber with pure \o2 to significantly raise the pressure above the nominal background and used an independent vacuum gauge to measure the absolute pressure. Since the RGA is mounted about 7 cm downstream from the trap we could measure only molecular beams that were not decelerated to very low velocities. Using a beam at 350 m/s we deduced using this procedure a peak beam density of $10^{10}$ molecules/cm$^3$. We were then able to infer the trapped molecules' density from the relative REMPI signal of the guided 350 m/s beam and trapped molecules.
The number of Li atoms here was too small to be detected with the RGA, hence we estimated it from the single-photon ionization cross section $\sigma_{Li}=1.6 \cdot 10^{-18}$ m$^2$ \cite{Hudson1967atomic}, which is much more reliable than REMPI. The beam shape here was also known to better accuracy, as in this case the atoms where not at the focus, but rather where the beam diverges. We estimate about $10^5$ trapped Li atoms, which indicates an inefficient entrainment.
            
In conclusion, by combining the high-intensity of a supersonic source with the efficiency of a moving trap decelerator we have demonstrated the largest ensemble of trapped cold molecules achieved to date.
We believe that a significant improvement will be gained by performing the ablation during the expansion stage, which requires modification of the nozzle geometry (work in progress).
Our method is applicable to other species, for example NH$_2$ and NH, where the latter has favorable Frank-Condon factors and in principle can be directly laser-cooled. In addition, magnetic deceleration is not limited to supersonic sources, and can be used together with other setups, such as buffer gas cooling systems. One of the exciting uses is in the direct cooling of molecules, such as SrF, which was recently trapped and cooled to temperatures as low as 2.5 mK, after being loaded to a MOT from a cryogenic buffer gas source and slowed using radiation pressure \cite{Norrgard2016Submillikelvin}.  This work paves the way for studying cold collisions of atoms and molecules in a magnetic trap, and for using sympathetic cooling in order to reach the ultra-cold regime for molecules at high densities.

\bibliographystyle{apsrev4-1}
\bibliography{O2LiPaper}

\end{document}